\documentclass[journal]{IEEEtran}
\usepackage{graphicx}
\usepackage{graphics}
\usepackage{caption}
\usepackage{algorithm}
\usepackage{algpseudocode}
\usepackage{balance}
\usepackage{amsmath}
\usepackage{bm}
\usepackage{url}
\usepackage{amssymb}
\usepackage{multirow}
\usepackage[table]{xcolor}
\usepackage{color}
\usepackage{cite}
\usepackage{caption}
\usepackage{subfigure}
\usepackage{color}
\usepackage{url}

\begin{document}

\title{Autonomous Vehicular Networks: Perspective and Open Issues}
\author{Tom H. Luan, Yao Zhang, Lin Cai, Yilong Hui, Changle Li, Nan Cheng
\thanks{Tom H. Luan, Yilong Hui, Changle Li and Nan Cheng are with Xidian University.}
\thanks{Yao Zhang is with The Hong Kong Polytechnic University.}
\thanks{Lin Cai is with University of Victoria.}
}
\maketitle

\begin{abstract}
The vehicular ad hoc networks (VANETs)\footnote{%
We consider VANETs as a general vehicle-to-vehicle communication framework
with the IEEE 802.11p or 5G CV2X wireless radio and Ad Hoc routing suite.}
have been researched for over twenty years. Although being a fundamental
communication approach for vehicles, the conventional VANETs are challenged
by the newly emerged autonomous vehicles (AVs) which introduce new features
and challenges on communications. In the meantime, with the recent advances
of artificial intelligence and 5G cellular networks, how should the
fundamental framework of VANET evolve to utilize the new technologies? In
this article, we reconsider the problem of vehicle-to-vehicle communications
when the network is composed of AVs. We discuss the features and specific
demands of AVs and how the conventional VANETs should adapt to fit them.
\end{abstract}



\IEEEpeerreviewmaketitle

\section{Introduction}

\label{sec: introduction}

It is no longer doubt that the autonomous vehicles are coming soon.
Nowadays, nearly all car manufactures like BMW, Volvo, Ford, have deployed
the semi-autonomous driving (a.k.a advanced driver-assistance system) on
their new models, which can provide the lane centering, adaptive cruise
control, autonomous emergency braking, \emph{etc}. As reported, all Tesla
vehicles shipped since 2019 have the full autonomous driving hardware, and
Tesla starts to offer the Full Self Driving 2.0 software package on their
new cars which enables autosteer in city streets. In December 2018, Waymo
launched the Waymo One self-driving taxi service, which allow users in the
Phoenix metropolitan area to request a pick-up service on their phones for a
fully autonomous driving minivan.

While the autonomous driving has gained substantial success world-wide, the
large-scale deployment of full autonomous driving on the city street is
still fraught with fundamental challenges. Notably, the current autonomous
vehicles mainly rely on the single-car intelligence, \emph{i.e.}, to sense
the environment and make driving decisions by the car itself. With the
complex and fast changing road environments, as well as the very limited
sensing and computing capability on board, the single-car intelligence not
only incurs expensive hardware to enable real-time sensing and artificial
intelligence (AI) processing, but also is prone to errors which may lead to
severe road accidents \cite{fu2020vehicular}.

\begin{figure}[t!]
\centering
\includegraphics[width=.85\linewidth]{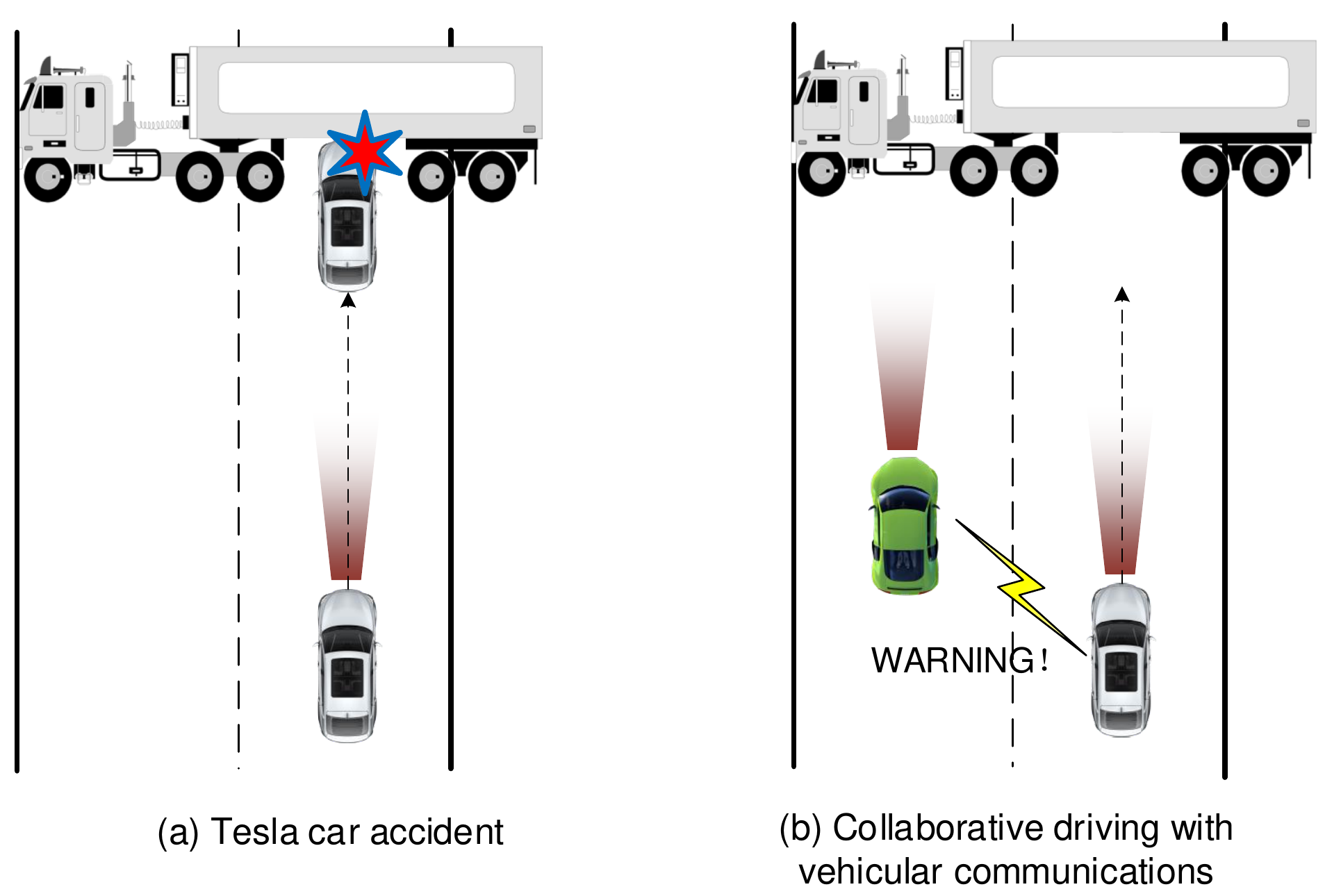}
\caption{Tesla car crash and cooperative driving. (a) The accident occurred
in May 16, 2016 when a truck was turning at an intersection on the divided
highway driving. (b) The vehicles on neighboring lane can detect the truck
and warn the other vehicles to avoid the crash.}
\label{fig: tesla}
\end{figure}

Fig.~\ref{fig: tesla}(a) illustrates a fatal Tesla car crash reported in
May, 2016. The accident was caused by the failure of multiple sensors at the
same time. Specifically, a Tesla Model S driving with Autopilot crashed into
the truck when the truck was turning at an intersection on the divided
highway. The body of the truck was white which made the camera and vision
system regarded the truck as a cloud. The radar signals traversed through
the gap between the tires, making the sensors considered the front to be the
entry of a tunnel. As a result, the car crashed into the hole of the track
without reducing speed.

The accident can be entirely avoided when vehicular communications and
collaborative driving come into play. As the example illustrated in Fig.~\ref%
{fig: tesla}(b), if there are any other vehicles on the neighboring lane and
can communicate with the accident car, they can detect the truck and warn
the accident car to avoid the fatal crash.

\begin{figure*}[t!]
\centering
\includegraphics[width=.85\linewidth]{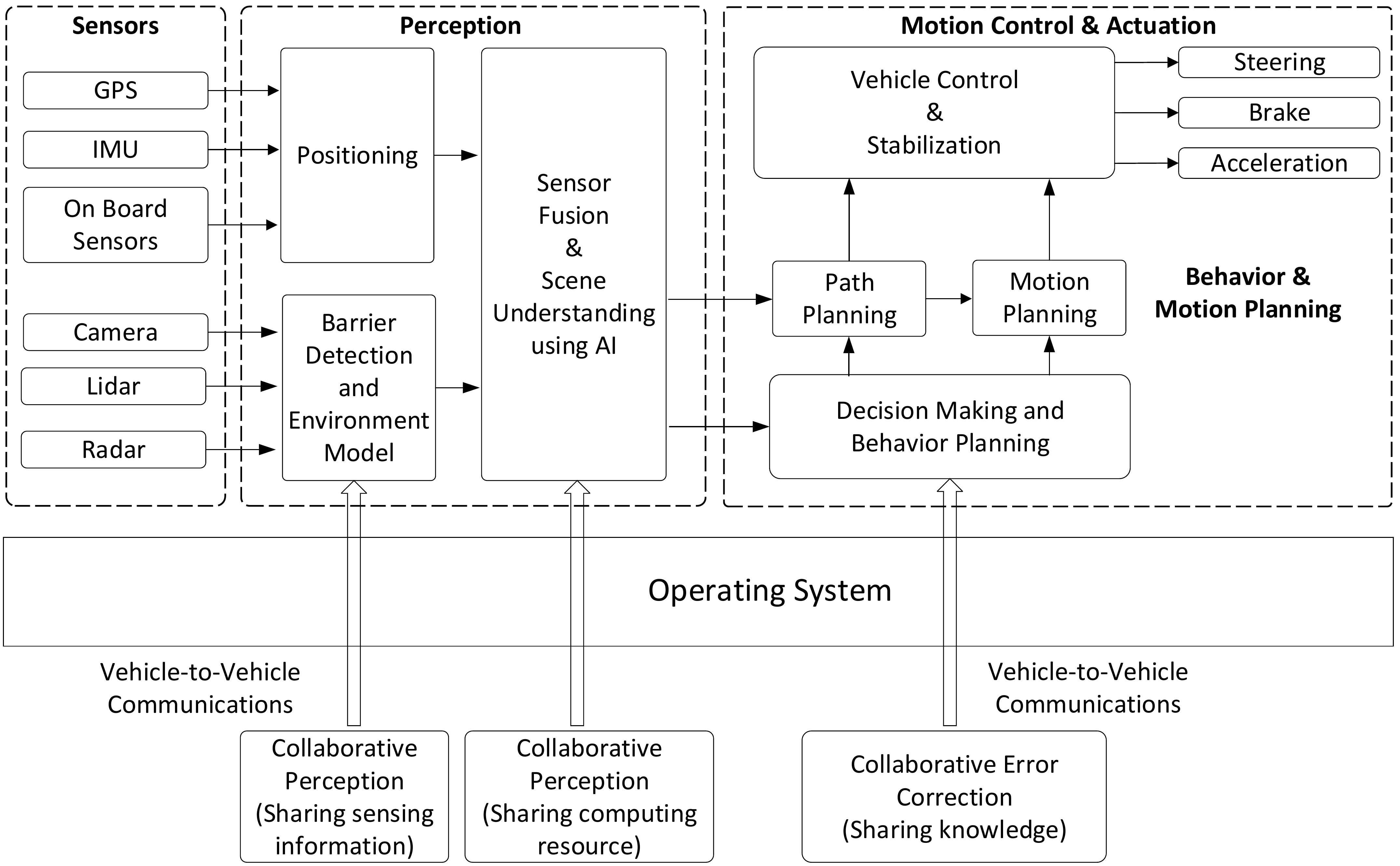}
\caption{Software architecture of an AV}
\label{fig: software}
\end{figure*}

The collaborative driving, by sharing the information among vehicles using
the inter-vehicular or vehicle-to-vehicle (V2V) communications\footnote{%
In this paper, we only consider the V2V communication mode. The
vehicle-to-infrastructure (V2I) communications however can also leverage the
technology discussed in the paper.}, can extend the single-car intelligence
to multi-car intelligence. The multi-car intelligence can significantly
reduce the computation load and cost on hardware compared with its
single-car counterpart, but also significantly improve the safety and
performance of driving. The multi-car intelligence can be achieved with the
following three aspects:

\begin{itemize}
\item \emph{Collaborative Sensing}: the neighboring vehicles can share the
sensing information among each other to not only extend the sensing range
but also enhance the sensing accuracy. The extended information achieved by
each vehicle can be crucial to make accurate driving decisions.

\item \emph{Collaborative Perception}: with real-time connections among each
other, the neighboring vehicles can perform the distributed machine
learning, \emph{e.g.}, meta-learning and federated learning, to
collaboratively learn the environment. The collaborative perception can
significantly reduce the computing burden of single vehicles and make more
reliable driving decisions.

\item \emph{Collaborative Error Correction}: the neighboring vehicles can
directly share the driving decisions of individuals to each other and help
others fix their error. As the example in Fig.~\ref{fig: tesla}(b), the
vehicle on the left lane can broadcast its \emph{break} decision and notify
the danger to the vehicle on the right lane to avoid the accident.
\end{itemize}

To enable the collaborative driving relies on the vehicular ad hoc networks
(VANETs) among AVs. Although been researched for decades, the conventional
VANETs mainly focus on delivering conventional status update or infotainment
messages, which have not considered much on the features of AVs and their
demand on collaborative computing and driving \cite{li2018building}. In
conventional VANETs, vehicles are pure mechanical systems without much
intelligence; the communications in VANETs thus do not consider any software
and computing advantages of vehicles. With the complete autonomous driving
and omni-AI capability on board, a new paradigm of autonomous vehicular ad
hoc networks (AVNETs), which can fully explore the features of AV and target
to enable the collaborative driving, needs to be structured.

In this article, we discuss the paradigm of AVNET and its key design
factors. We first describe the software architecture of an AV, which is
important to the design of the communication systems. After that, we compare
the autonomous vehicular networks with the conventional vehicular ad hoc
networks and identify the new features of AVNET. We then showcase an example
by applying the mmWave V2V communications in AVNET and evaluate the
performance using simulations. Lastly, we conclude the article with the
discussion on open research issues.

\section{Software Architecture of AV}

\label{sec: software}

Fig.~\ref{fig: software} shows the software architecture of an AV, which can
be divided into three parts: sensing, perception and motion control.

\subsection{Sensing}

An AV is typically equipped with rich sensors for localization and
environment sensing, including the LiDAR, ultrasonic radar, mmWave radar,
camera, GPS, GNSS, \emph{etc.} Fig.~\ref{fig: software} shows the primary
sensors used in the AV and their key features. For example, Waymo has 29
cameras integrated around the body of the car, a LiDAR providing higher
resolution across a 360 degree field of view with over 300-meter range, and
radars to enhance the accuracy of sensing in extreme weather conditions,
\emph{e.g.}, rain, fog and snow. The Tesla cars deploy eight surround
cameras, twelve ultrasonic sensors and a forward-facing mmWave radar without
the LiDAR to reduce the hardware cost. A high resolution camera typically
produces data at the rate of $500-11,500$ Mbit/s whereas a LiDAR produces
data streams ranging from $20- 100$ Mbit/s. As reported, a typical test
vehicle equipped with $3-4$ cameras, $3-4$ LiDAR sensors, and other sensors
create up to $10-20$ TB per hour.

\subsection{Perception}

Perception is the data processing module of AVs, which is used to understand
the driving environments and output important perception information.

While there exists many kinds of perception systems, the core pipeline
typically includes object detection, feature extraction, motion estimation,
object tracking, mapping and localization. We roughly divide the current
advances of computer vision technologies in perception systems into two
types, \emph{i.e.}, object-aware methods and feature-aware methods. For the
first one, two methods are popular and have potential in vehicular
perception, \emph{i.e.}, labelling each bounding box at frames (such YOLO
\cite{du2020server}) and labelling each pixel at frames (semantic
segmentation, such as Mask RCNN \cite{du2020server}).

Different from object-aware CV technologies, Simultaneous Localization and
Mapping (SLAM) systems focus on feature-level data processing, in order to
support the real-time mapping and localization of vehicles \cite{mur2017orb}%
. SLAM provides a suite of cost-effective logics to collect data from camera
or LiDAR and then output HD map and localization results. Therefore,
combining the advantages of object detection (or semantic segmentation) and
SLAM, \emph{i.e.}, using SLAM pipelines with the transferred object
detection neural network, is a new way to build vehicular perception system.

Using the inter-vehicle collaboration over the existing perception system,
not only the CV related performance metrics, such as detection accuracy,
motion estimation error, could be raised, but also resources of computing
and communication will be saved. The insight behind the resource saving is
the adaptation of system configurations such as sensing views, frames per
second, frame resolution selection, \emph{etc.} Thus the configuration
adaptation leaves space for the exploitation of collaborative perception
(sharing sensing information) and collaborative perception (sharing AI
networks).

\subsection{Motion Control and Actuation}

The ultimate goal of sensing and perception is to determine the optimal
behavior and motion control of AVs. As in Fig.~\ref{fig: software}, the
motion control relies on not only the fusion of sensing and perception, but
also the information shared by neighboring vehicles using the collaborative
error correction \cite{fu2019graded}. The collective information is
processed by machine learning algorithms towards the optimal maneuver
control of vehicles.

\section{Vehicular Communication Technology}

\label{sec: VCT}

Multi-car intelligence relies on real-time wireless communications among
vehicles. The vehicular communications have been standardized by IEEE and
Third Generation Partnership Project (3GPP), respectively.

\subsection{Dedicated Short-Range Communication Technology (DSRC)}

In 1999, the US Federal Communications Commission (FCC) has allocated a
spectrum of 75 MHz at the frequency band of 5.9 Hz for the dedicated short
range communication (DSRC) for vehicles. The DSRC adopts the IEEE 802.11p
Wireless Access in Vehicular Environments (WAVE) standard which reuses the
IEEE 802.11g OFDM physical layer and IEEE 802.11e Enhanced Distributed
Channel Access (EDCA) MAC layer. The EDCA MAC is contention-based which
provides differentiated priorities to safety and infotainment traffics.
While being extensively researched for over twenty years, the DSRC has not
been widely deployed. This is because that while the IEEE 802.11 protocol
achieves great success for static indoor applications, the short-range
contention-based scheme does not fit the usage in a highly dynamic and
possibly populated outdoor vehicular environment for applications related to
public safety \cite{luan2011mac}.

\subsection{Cellular-based Vehicular Network (C-V2X)}

C-V2X is defined by 3GPP Release 14 in 2017 which adopts the cellular
technology for vehicular communications \cite{chen2020vision}. The C-V2X
exploits the LTE uplink and utilizes Single Carrier Frequency Division
Multiple Access (SC-FDMA) at the physical and MAC layers. The
vehicle-to-vehicle communications in C-V2X adopt the LTE sidelink technology
which operates in a distributed manner without the need of cellular
infrastructure. As compared to DSRC, the C-V2X uses the cellular time and
frequency division based MAC with reserved resource blocks for each
transmission which is more reliable.

\subsection{mmWave Communications}

To meet the futuristic application needs of vehicular communications, both
IEEE and 3GPP have made new amendments on their V2V standards. The 802.11
standard group has defined IEEE 802.11bd protocol to replace IEEE 802.11p,
and 3GPP defines the next generation of C-V2X with New Radio V2X (NR-V2X) in
Release 16 in June 2019.

Both IEEE 802.11bd and 5G NR V2X standards will support operations at mmWave
frequencies, \emph{i.e.}, in the 57-71 GHz and 24.25-52.6 GHz ranges,
respectively. The large spectrum availability at these frequencies promotes
high-capacity low-latency communication. More importantly, since mmWave is
directional in nature, it is desirable for vehicular communications where
vehicles are constrained by roads.


\section{Platform Bridging the Computation and Communications}

\label{sec: os}

Orchestrating the complicated task processing in vehicular perception
systems and real-time communications of vehicles needs a platform. The Robot
Operating System (ROS) gives an important insight.

Many successful AV operating systems, \emph{e.g.}, Baidu Apollo,
autoware.ai, root from ROS, an open-source software suite for robotics
application development. ROS provides the necessary operating system
application services to manage a robotic system, such as robot models,
perception, location and mapping, path planning, navigation, simulation
tools, \emph{etc}.

\begin{figure}[tbp]
\centering
\includegraphics[width=0.8\linewidth]{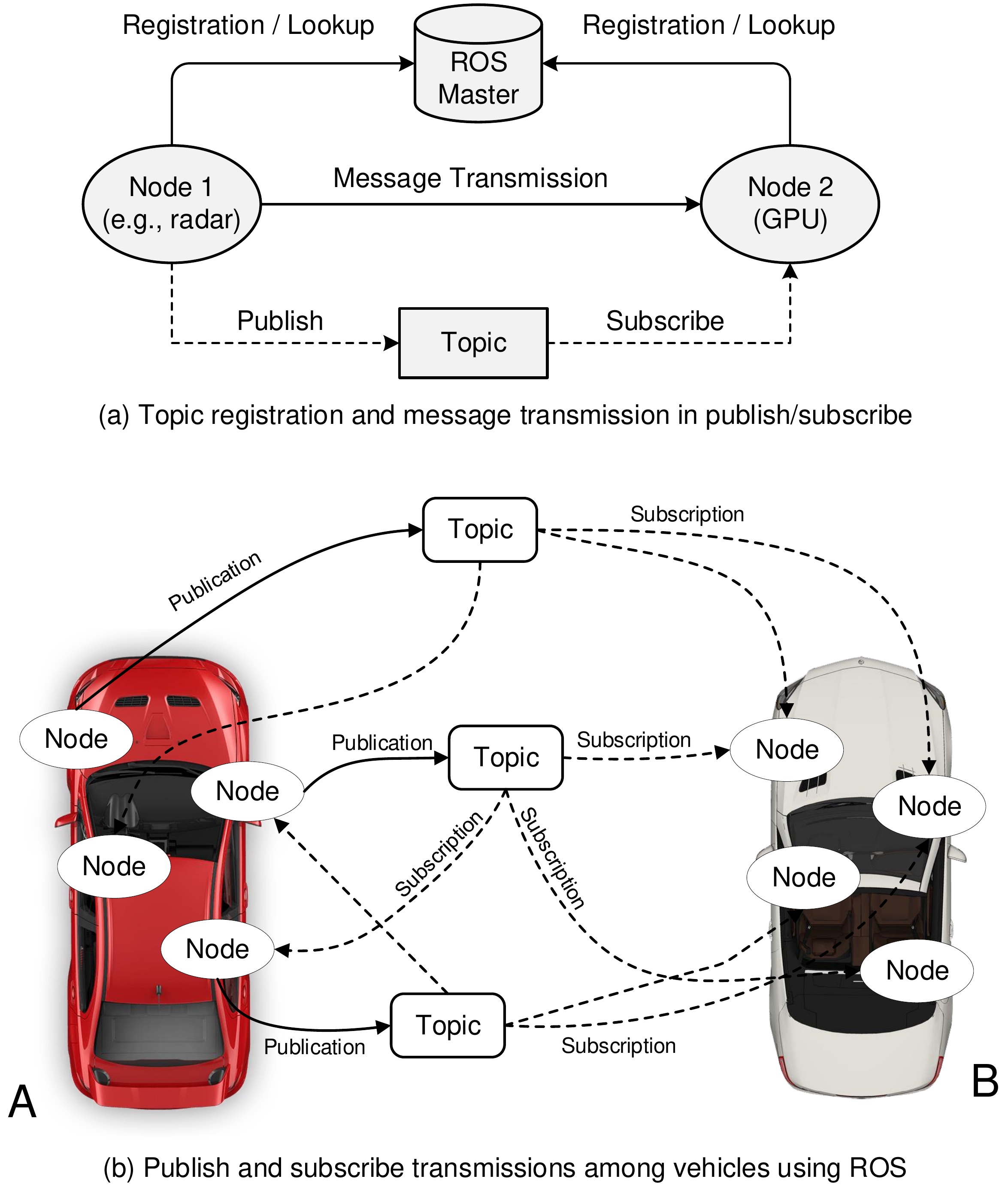}
\caption{Publish/subscribe mechanism in ROS}
\label{fig: ros}
\end{figure}

It is important to note that in order to support the real-time data
communications between the sensing and computing units, ROS has defined its
communication system based on the publish/subscribe (short as pub/sub)
scheme. Specifically, as shown in Fig.~\ref{fig: ros}(a), ROS considers each
component of data to transmit or receive, \emph{e.g.}, radar, camera, CPU,
as a participant. A ''global data space", ROS Master in Fig.~\ref{fig: ros}%
(a), is maintained which can be accessed by all independent applications.
The data transmissions among participants are accomplished via topics. For
instance, a component (\emph{e.g.}, a camera) publishes a topic (\emph{e.g.}%
, forward image) in the ''global data space", and the components who would
be interested in the data subscribes to the topic. When new data are
collected, the publisher would publish the updated data and all subscribers
of the topic can retrieve the data.

The pub/sub \cite{eugster2003many} scheme builds a bridge between the task
processing of the publisher (the current vehicle) and active perception
pulling from the participants (nearby vehicles). The bridge shifts the
conventional VANET pattern to active pull-based information sharing.
Particularly, the pub/sub scheme has following three features, making it
very suitable to data sharing in collaborative driving.

\begin{itemize}
\item \emph{Pull-based approach}: in the pub/sub scheme, the components
select to pull the data from the interested topics based on their own
requirements, whereas the data publishers do not worry about where to send
the data. This is reasonable as in an AV, the publishers which are typically
sensors have limited processing power and cannot perform complicated
operations to determine where to send their data. On the other hand, in a
dynamic environment, a participant may need to dynamically change topics of
data to download. The pull-based approach is more reliable for a dynamic
autonomous system.

\item \emph{Fully distributed system}: the pub/sub scheme has been used for
both intra-vehicular and inter-vehicular communications. As plotted in Fig.~%
\ref{fig: ros}(b), all the components of a vehicle are distributed and peer
participants. The components of other vehicles, \emph{e.g.}, car B, are also
participants which can subscribe to the topics. As a result, the GPU of car
B can subscribe to the car A's camera topic, and download the data when car
A's camera publishes new data. In this manner, collaborative perception is
enabled.

\item \emph{Multicast}: the pub/sub scheme is a multicast system by nature.
A topic can be subscribed by multiple subscribers intra a vehicle or inter
vehicles at the same time. The update data of the topic will be accordingly
multicast to all subscribers. Therefore, the AVNETs are mainly composed of
multicast traffic flows managed by the topics which could be on sensing,
computing model or computing results.
\end{itemize}

\section{Autonomous Vehicular Networks v.s. Conventional Vehicular Ad Hoc
Networks}

In this section, we present three key features of the AVNETs distinguished
from the conventional VANETs.

\subsection{Link: Blind v.s. Sighted Communications}

The conventional vehicles in VANET is a mechanism system without the
capability of environment perception. As a result, the vehicles in VANET are
almost ''blind" and send the information using omn-directional antenna.
Nevertheless, the AVs are of full capability to ''perceive" the environment
using various sensors, LiDAR and cameras. Therefore, AVs can gauge their
communication beams towards the best physical layer performance. \cite%
{mashhadi2021federated} develops a federated learning framework which use
the LiDAR system to assist mmWave beamforming for vehicle-to-infrastructure
communications.

\begin{figure*}[t!]
\centering
\includegraphics[width=0.9\textwidth]{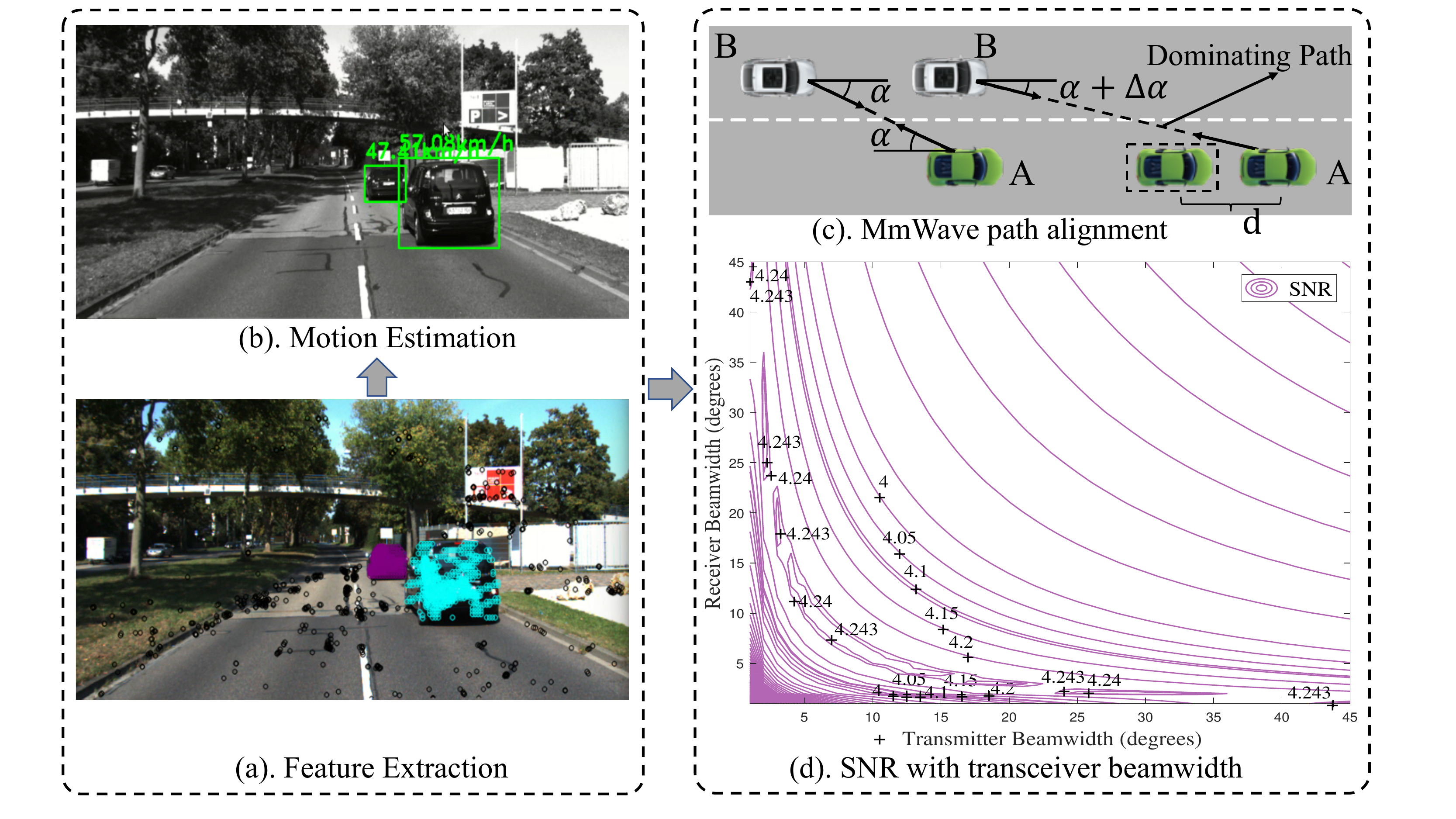}
\caption{From Sensing to Communications}
\label{SensingtoCom}
\end{figure*}

Fig.~\ref{SensingtoCom}(a)-(b) show the processed sensor data viewed by the
current vehicle. With the capability to ''see" the targets for
communications, the AVs can set up an optimized mmWave beam for
communications and adapts to the vehicle's mobility with the assistance of
onboard perception devices.

\subsection{Topology: Self-Controlled v.s. Human Controlled Topology}

The topology among nodes plays an important role in the communication
performance; the distance and relative velocity among vehicles determine the
physical layer transmission capacity and the MAC contentions. Traditional
vehicles in VANETs are controlled by human beings, and the communications
among vehicles mainly serves to deliver the road information, \emph{e.g.},
safety messages \cite{bi2015multi}, to assist driving. However, in AVNETs,
the nodes are self-driving vehicles, which can fully control their own
driving velocity and relative locations to each other, as long as their
passengers can reach their transportation goals safely and comfortably. As a
result, to benefit the collaborative sensing and computing, it is entirely
possible for vehicles to collaboratively adapt their topology to achieve
better transportation, sensing and communication performance.

Notably, there is a golden rule in VANETs that the communications cannot
directly change the driving of drivers. This is no longer a restriction in
AVNET. The new dimension would significantly benefit the communications.

\begin{table}[tbp]
\caption{Comparisons between VANET and AVNET}\centering%
\begin{tabular}{c||c|c|c}
& Communication & Routing & Mobility \\ \hline\hline
& IEEE 802.11p & Ad Hoc &  \\
VANET & (omni-directional & (e.g., OLSR, & Human \\
& transmission) & AODV) &  \\ \hline
& IEEE 802.11bd, &  &  \\
AVNET &  5G NR V2X & Publish/Subscribe, & AI \\
& (directional transmission) & Ad Hoc &  \\ \hline
\end{tabular}%
\label{tab: avn}
\end{table}

\subsection{Routing: Explicit IP Routing v.s. Implicit Published/Subscribe
Routing}

The VANET typically adopts the Ad Hoc approaches for traffic routing, which
rely on the Internet layer 2 or layer 3 addressing approach to route the
traffic \cite{8354811}. The routing methods are based on explicit
addressing, where the traffic are routed to receivers with explicit IP or
MAC addresses.

Unlike the explicit addressing, as discussed in Section \ref{sec: os},
AVNETs are prone to use the implicit addressing approach such as the pub/sub
mechanism for routing traffic, as specified in ROS; the implicit addressing%
\footnote{%
Another example of the implicit routing is the named data networking (NDN)
which uses ''interests" to route data traffic.} uses ''topics" for routing,
instead of the explicit address. This is because that AVNET is mainly used
to route collaborative sensing and computing information. The pub/sub
mechanism is a multicast protocol in nature and a pull-based approach which
is more suitable to transmit commonly interested sensing and computation
traffic among vehicles.

TABLE~\ref{tab: avn} summarizes the comparison between AVNETs and VANETs
from the perspectively of radio links, traffic routing and network topology.

\section{Use Case of AVNET} \label{sec: use case}

In this section, we provide a simulation study on the collaborative AVNET.

\subsection{Framework}

The interaction between V2V mmWave communications and vehicular sensing is
depicted. Specifically, two vehicles are assumed to form a simple platoon.
They collect perception information from environments and communicate with
each other using mmWave links. Our purpose is to evaluate the V2V mmWave
performance with different bandwidth so that reflect the advantage of mmWave
technologies in vehicular scenarios. To utilize the high-throughput mmWave
links, reliable beam alignment is necessary, especially for two moving
vehicles. To do that, we develop a system to speedup the beam alignment
process using some helpful information extracted from vehicular sensing
components. Therefore, there are two components in our performance
evaluation system, vehicular sensing and beam alignment. In our system, the
frequent feedback used for beam realignment between two transceivers can be
eliminated. On the other hand, the high throughput of V2V mmWave links can
also help the perception sharing between vehicles, which further help
vehicles understand driving environments efficiently.

\subsection{Performance Evaluation}

Fig.~\ref{SensingtoCom} shows a pipeline of the performance evaluation
system that combines the sensing and inter-vehicle communications, which
showcases the positive effect of sensing on V2V communications. The sensor
data processing components are achieved by the basic SLAM system with object
detection capability \cite{zhang2020vdo}. The detected objects and extracted
object features are shown in Fig.~\ref{SensingtoCom} (a). Using the temporal
frame dynamics, the motion information of adjacent vehicles can be estimated
so that the velocity and relative angle could be acquired by the current
vehicle, as shown in Fig.~\ref{SensingtoCom} (b). The motion estimation
results are then used to guide the beamforming of inter-vehicle mmWave link,
as shown in Fig.~\ref{SensingtoCom} (c). In Fig.~\ref{SensingtoCom} (d), we
then present the SNR performance impacted by the beamwidth. Fig.~\ref%
{SensingtoCom} gives an important insight that how to develop inter-vehicle
mmWave links more efficiently using sensing information rather than the
traditional beamforming methods.

\begin{figure}[ht!]
\centering
\includegraphics[width=\linewidth]{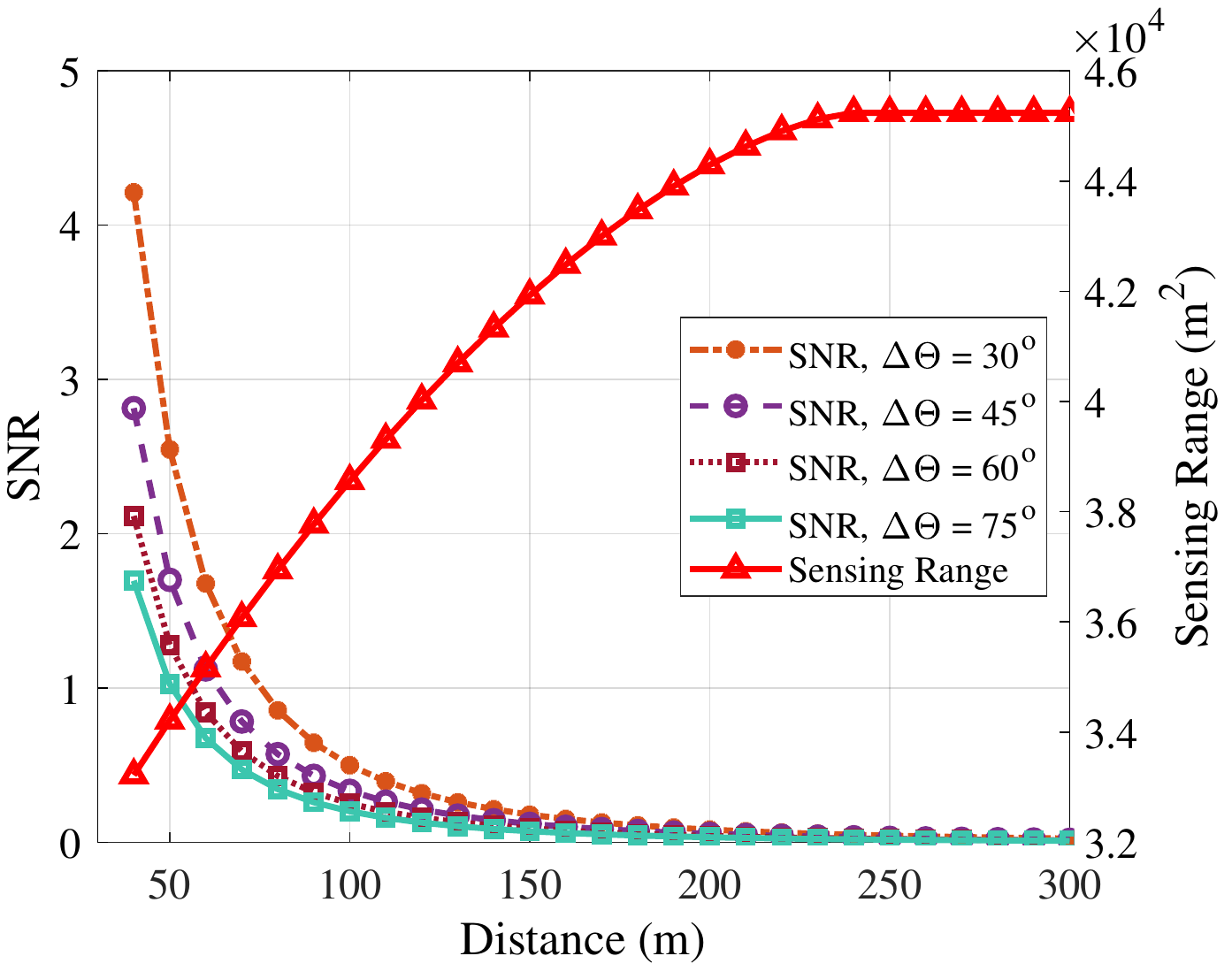}
\caption{Performance of Platoon Cooperative Sensing}
\label{sensingVSsnr}
\end{figure}

To show the communication performance of mmWave empowered platooning, two
metrics are further discussed, \emph{i.e.}, directional mmWave link
performance and cooperative sensing range, as well as their potential
trade-off with the changing V2V distance. All results are shown in Fig.~\ref%
{sensingVSsnr}. The first metric is SNR (Signal-to-Noise-Ratio), showing the
impact of V2V distance on beamwidth. We set the relative beamwidth between
transmitter and receiver as four different values. For simplicity, the
receiver SNR only includes the directive gains in the main lobe of mmWave
V2V links by resorting to the channel model and antenna pattern in \cite%
{perfecto2017millimeter}. 
We also depict the sensing range impacted by V2V distance in Fig.~\ref%
{sensingVSsnr}. According to \cite{qiu2018avr}, the sensing range is an
important metric in existing commercial on-board perception devices. For
example, a 64-bead LiDAR can collect 2.2 million points each second and
achieve real-time sensing of $360^{\circ}$ view \cite{qiu2018avr}. Assuming
that each vehicle achieves a full-view sensing, the sensing range of
vehicles in a platoon thus may overlay each other.
The platoon provides a new cooperative sensing paradigm where two nearby
vehicles can collaborate with each other and extend their individual sensing
range. Considering two vehicles with full-view sensing sensors, the
aggregated sensing range that reflects the cooperative perception capability
of two platoon vehicles is evaluated in Fig.~\ref{sensingVSsnr}.

We select those two metrics to show that the platoon based AVNET can combine
the advantages of mmWave vehicular communications and on-board sensing,
which can enhance the performance of platoon based autonomous driving. From
Fig.~\ref{sensingVSsnr} we can also see that, by combining mmWave and
sensing in AVNET, many specific technical problems will motivate researchers
to develop new algorithms, models or architectures to balance the
performance of vehicular sensing and V2V transmissions.


\section{Open Research Issues}

\label{section:conclusion}

\emph{AI Enabled Communications and Networking}: with the AVs of strong
software and hardware computing and communication capabilities as the main
entities of communications, how to fully explore the omni-AI of AVs towards
the optimal application performance deserve in-depth study. In this article,
we have demonstrated three dimensions for exploration. However, how to fully
explore the new dimensions and apply the advanced AI algorithms to optimize
the performance from the shoe of the entire system remain open for research.

\emph{AVs Empowered Intelligent Transforation System (ITS)}: the ITS relies
on advanced information system and wireless communication technologies to
connect the traffic control units with the transportation units. As a
result, the traffic can be intelligently coordinated towards the best
transportation performance from the perspective of traffic efficiency,
environment and social utility. The traditional ITS is based on VANETs in
which the vehicles are operated by human beings. With AVs widely adopted in
the future, the ITS can significantly benefit from the controllable driving
of AVs towards a global optimal transportation performance. How to manage
the city-wide AV in ITS is open for research \cite{hui2021unmanned}.

\emph{Security}: different from the traditional vehicles controlled by
human, the AVs require the timely processing of a titanic amount of data for
driving decision making. As a result, collaborative driving with
collaborative sensing, perception and error correction becomes important in
AVNETs. The collaborations among AVs however impose significant security
challenges. With the features of sighted communications, pub/sub and
collaborative topology control mechanisms demonstrated, new security
threatens to each feature may raise and deserve further study.

\section{Conclusion}

The research of VANETs has been carried out for more than 20 years, and a
large number of outstanding research results have been derived. However,
legacy VANETs only consider traditional mechanical vehicles as the main
communication subjects, and have not considered the intelligence of
vehicles. Emerging AVs have comprehensive environmental awareness
capabilities, as well as strong intelligence and control on driving and
wireless transmission strategies. The adoption of AVs therefore requires a
revisit on the design of VANETs to not only accommodate the new service
requirements of AVs, but also fully explores the AI features of AVs. In this
article, we have discussed the software structure and operating system of an
AV, and demonstrated the three features of AVNETs from the perspective of
wireless transmission, traffic routing and topology control. Through
simulation research, we demonstrated the performance of collaborative
driving using AVNET.

\ifCLASSOPTIONcaptionsoff
\newpage \fi

\bibliographystyle{IEEEtran}
\bibliography{AVN}

\begin{thebibliography}{10}
\providecommand{\url}[1]{#1}
\csname url@samestyle\endcsname
\providecommand{\newblock}{\relax}
\providecommand{\bibinfo}[2]{#2}
\providecommand{\BIBentrySTDinterwordspacing}{\spaceskip=0pt\relax}
\providecommand{\BIBentryALTinterwordstretchfactor}{4}
\providecommand{\BIBentryALTinterwordspacing}{\spaceskip=\fontdimen2\font plus
\BIBentryALTinterwordstretchfactor\fontdimen3\font minus
  \fontdimen4\font\relax}
\providecommand{\BIBforeignlanguage}[2]{{%
\expandafter\ifx\csname l@#1\endcsname\relax
\typeout{** WARNING: IEEEtran.bst: No hyphenation pattern has been}%
\typeout{** loaded for the language `#1'. Using the pattern for}%
\typeout{** the default language instead.}%
\else
\language=\csname l@#1\endcsname
\fi
#2}}
\providecommand{\BIBdecl}{\relax}
\BIBdecl

\bibitem{fu2020vehicular}
Y.~Fu, F.~R. Yu, C.~Li, T.~H. Luan, and Y.~Zhang, ``Vehicular blockchain-based
  collective learning for connected and autonomous vehicles,'' \emph{IEEE
  Wireless Communications}, vol.~27, no.~2, pp. 197--203, 2020.

\bibitem{li2018building}
C.~Li, Y.~Zhang, T.~H. Luan, and Y.~Fu, ``Building transmission backbone for
  highway vehicular networks: Framework and analysis,'' \emph{IEEE Transactions
  on Vehicular Technology}, vol.~67, no.~9, pp. 8709--8722, 2018.

\bibitem{du2020server}
K.~Du, A.~Pervaiz, X.~Yuan, A.~Chowdhery, Q.~Zhang, H.~Hoffmann, and J.~Jiang,
  ``Server-driven video streaming for deep learning inference,'' in
  \emph{Proceedings of ACM Sigcomm}, 2020, pp. 557--570.

\bibitem{mur2017orb}
R.~Mur-Artal and J.~D. Tard{\'o}s, ``{Orb-slam2: An open-source SLAM system for
  monocular, stereo, and rgb-d cameras},'' \emph{IEEE Transactions on
  Robotics}, vol.~33, no.~5, pp. 1255--1262, 2017.

\bibitem{fu2019graded}
Y.~Fu, C.~Li, T.~H. Luan, Y.~Zhang, and F.~R. Yu, ``Graded warning for rear-end
  collision: An artificial intelligence-aided algorithm,'' \emph{IEEE
  Transactions on Intelligent Transportation Systems}, vol.~21, no.~2, pp.
  565--579, 2019.

\bibitem{luan2011mac}
T.~H. Luan, X.~Ling, and X.~Shen, ``{MAC in motion: Impact of mobility on the
  MAC of drive-thru Internet},'' \emph{IEEE Transactions on Mobile Computing},
  vol.~11, no.~2, pp. 305--319, 2011.

\bibitem{chen2020vision}
S.~Chen, J.~Hu, Y.~Shi, L.~Zhao, and W.~Li, ``A vision of {C-V2X}:
  technologies, field testing, and challenges with chinese development,''
  \emph{IEEE Internet of Things Journal}, vol.~7, no.~5, pp. 3872--3881, 2020.

\bibitem{eugster2003many}
P.~T. Eugster, P.~A. Felber, R.~Guerraoui, and A.-M. Kermarrec, ``The many
  faces of publish/subscribe,'' \emph{ACM computing surveys}, vol.~35, no.~2,
  pp. 114--131, 2003.

\bibitem{mashhadi2021federated}
M.~B. Mashhadi, M.~Jankowski, T.-Y. Tung, S.~Kobus, and D.~Gunduz, ``Federated
  mmwave beam selection utilizing lidar data,'' \emph{arXiv preprint
  arXiv:2102.02802}, 2021.

\bibitem{bi2015multi}
Y.~Bi, H.~Shan, X.~S. Shen, N.~Wang, and H.~Zhao, ``A multi-hop broadcast
  protocol for emergency message dissemination in urban vehicular ad hoc
  networks,'' \emph{IEEE Transactions on Intelligent Transportation Systems},
  vol.~17, no.~3, pp. 736--750, 2015.

\bibitem{8354811}
H.~Peng, L.~Liang, X.~Shen, and G.~Y. Li, ``Vehicular communications: A network
  layer perspective,'' \emph{IEEE Transactions on Vehicular Technology},
  vol.~68, no.~2, pp. 1064--1078, 2019.

\bibitem{zhang2020vdo}
J.~Zhang, M.~Henein, R.~Mahony, and V.~Ila, ``{VDO-SLAM}: A visual dynamic
  object-aware slam system,'' \emph{arXiv preprint arXiv:2005.11052}, 2020.

\bibitem{perfecto2017millimeter}
C.~Perfecto, J.~Del~Ser, and M.~Bennis, ``Millimeter-wave {V2V} communications:
  Distributed association and beam alignment,'' \emph{IEEE Journal on Selected
  Areas in Communications}, vol.~35, no.~9, pp. 2148--2162, 2017.

\bibitem{qiu2018avr}
H.~Qiu, F.~Ahmad, F.~Bai, M.~Gruteser, and R.~Govindan, ``{AVR: Augmented
  vehicular reality},'' in \emph{Proceedings of ACM Mobisys}, 2018, pp. 81--95.

\bibitem{hui2021unmanned}
Y.~Hui, Z.~Su, and T.~H. Luan, ``Unmanned era: A service response framework in
  smart city,'' \emph{IEEE Transactions on Intelligent Transportation Systems},
  2021.

\end{thebibliography}

\end{document}